\documentclass[intlimits,twoside,a4paper]{article}

\usepackage[active]{srcltx}

\usepackage{amsmath,amssymb}
\usepackage{graphicx}
\usepackage{wrapfig}
\usepackage{xcolor}
\usepackage{amsfonts}
\usepackage{amsmath}
\usepackage[T2A]{fontenc}
\usepackage[cp1251]{inputenc}
\usepackage{cmpj2}

\issue{2016}{19}{3}{33602}
\doinumber{10.5488/CMP.19.33602}

\title[Bilayers of Janus-like particles]
{First-order phase transitions in lattice bilayers of Janus-like particles:
Monte Carlo simulations}

\author[M.~Bor\'owko \textsl{et al.}]
{M.~Bor\'owko, A.~Patrykiejew, W.~R\.zysko, S.~Soko\l owski}
\address{
Department for the Modelling of Physico-Chemical Processes,
Maria Curie-Sk\l odowska University, \\ Gliniana 33, Lublin, Poland}

\authorcopyright{M.~Bor\'owko, A.~Patrykiejew, W.~R\.zysko, S.~Soko\l owski, 2016}
\date{Received September 28, 2015, in final form February 2, 2016}

\begin{document}

\maketitle

\begin{abstract}
The first-order phase transitions in the lattice model of Janus-like particles confined in
slit-like pores are studied. We assume a cubic lattice with molecules that can freely
change their orientation on a lattice site. Moreover, the molecules can interact
with the pore walls with orientation-dependent forces. The performed calculations
are limited to the cases of bilayers.
Our emphasis is on the competition between the fluid-wall and fluid-fluid
interactions.
The oriented structures formed in the systems
in which the fluid-wall interactions acting contrary to the
fluid-fluid interactions differ from those appearing in the systems with
neutral walls or with walls attracting the repulsive parts of fluid molecules.


\end{abstract}

\section{Introduction}
A growing interest has been recently observed regarding the self-assembly of Janus and patchy colloids
into differently ordered structures. This is also due to the
progress in the development of new experimental methods for obtaining particles that are capable of building blocks
of the requested symmetry and structure \cite{1,2,3,4,5a,5,5b}.
One way to prepare spherically symmetric particles with anisotropic
interactions relies on the creation of attractive patches on their
surfaces \cite{4,6}. Advanced experimental methods allow for the
production of patchy colloids with different sizes, shapes and
chemical compositions \cite{8,9,10,11,12,13}.
In parallel with  experiments,
remarkable theoretical and computer simulation efforts have been undertaken to
predict the physical properties of these systems \cite{14}.

An important direction of the research  is connected with investigations
of phase transitions in such systems. The directional dependence
of interactions  leads to a very rich phase behavior and to the development
of various mesoscale structures \cite{15,16,17,18,19,20,21,22,23,24,25,26}.
A review of unusual structures predicted for
several types of colloids can be found in \cite{2}.
In particular,
in the case of patchy spheres  interacting via
the Kern-Frenkel potential \cite{27}, diverse aggregates,
 such as micelles, vesicles, and
bilayers, have been found \cite{20,21,22}.
Also, the coexistence
of a dilute phase of micelles and denser phase of larger
clusters has been observed. More recently \cite{22}  simulations of
 the formation of different crystalline structures have been carried out.
Another interesting kind of behavior has been obtained within the framework
of Wertheim theory \cite{28} for the patchy colloids with multiple sites,
which, in some cases, makes multiple bonding possible between sites \cite{34,35,36,37,38,39}.

 Lattice models have also been used to study the self-assembly. For example,
Dawson and Kurtovi{\'c} \cite{40}   introduced a  lattice
model of amphiphilic self-assembly in mixtures containing
amphiphiles and water. Davis and Panagiotopoulos \cite{41}
examined mixtures of flexible
chains comprising neutral and solvophobic monomers and
big amphiphiles with multiple chains connected to them
at  common attachment points. They found
that amphiphile geometry plays a key role in determining whether the micelles  are spheres or flat bilayers.
The behavior of Janus particles near solid surfaces was also investigated using density
functional theory and computer simulations \cite{41a,41b}.

Recently, two of us reported the results of Monte Carlo
simulations of the phase behavior of Janus disks on a square
lattice \cite{42}. The particles were composed of two different
parts and interactions between neighboring particles depended on
their orientations. It was found that the systems exhibited the
first-order transitions between colloidal-rich and colloidal-poor
phases and continuous order-disorder transitions to differently
ordered structures.

Calculations performed in \cite{42}  assumed a restricted number
of possible orientations of the particles. In this work we report
Monte Carlo studies of bilayers of Janus-like molecules on a cubic
lattice confined in slits, but in contrast to the previous work
\cite{42}, the molecules can freely change their orientations on
the sites. Moreover, the slit walls can also interact with
particles with orientation-dependent potential. Our principal aim
is to study the first-order transitions in the system and the
dependence of the phase behavior on the competition between
interparticle and the particle-wall interactions. The principal
order-parameter is density. However, we neither carried out
systematic studies of the finite size effects \cite{43aa,44aa} nor
the studies of  continuous transitions in the system.

\section{Simulation}
We consider a cubic lattice of the
size $N_x\times N_x\times N_z\,$,
$N_z= 2$ in $X$, $Y$ and $Z$ directions, respectively.
The lattice constant is assumed to be unit of length.
Periodic boundary conditions are used in the $X$ and $Y$ directions. In the $Z$
direction, however, the system is closed by planar walls, located at $i_z=0$ (the bottom
wall) and at $i_z=3$ (the top wall). We consider spherical particles consisting of two hemispheres,
one being solvophobic (A) and the other one being solvophilic (R). The solvent is involved implicitly.
We introduce the
occupation number for each lattice site, $n_{\bf i}$, ${\bf i}=(i_x,\,i_y,\,i_z)$ that equals 0 for an
empty site and 1 for an occupied site. The state of a
molecule  occupying  the
 site ${\bf i}$
is characterized by a unit vector $(\sin\theta_{\bf i}\cos\phi_{\bf i},\, \sin\theta_{\bf i}\sin\phi_{\bf i},\, \cos\theta_{\bf i})$,
where $\theta_{\bf i}$ is the tilting angle measured with respect to the vector
perpendicular to the bottom wall and $\phi_{\bf i}$ is the azimuthal angle, measured
with respect to the positive $X$ axis.

Due to the existing walls, the
energy experienced by a molecule  located on the site ${\bf i}$ is given by Yukawa-like
potential
\begin{equation}
 v_{\bf i}=\varepsilon_{\text{s}}\left[f(i_z)-f(N_z+1-i_z)\right]\cos\theta_{\bf i},
\end{equation}
where $f(i_z)=\exp(-i_z)/i_z\,$.

To evaluate the interparticle potential energy, we take into account only
the nearest-neighbor interactions. If two nearest-neighbor
sites, ${\bf i}$ and ${\bf j}$, are occupied,
the pair potential energy is
\begin{equation}
 u_{\bf ij}=\varepsilon (\cos\alpha_{\bf i} - \cos\alpha_{\bf j}),
\end{equation}
where $\alpha_{\bf i}$ and $\alpha_{\bf j}$ are the angles
measured with respect to the unit vector pointing from the site
${\bf j}$ to the site~${\bf i}$. A similar potential of
anisotropic interactions  was used for amphiphilic molecules in
\cite{41a}. We assume that  $\varepsilon>0$. In the absence of an
external field, the particles prefer to couple in pairs whose
A-patches face each other. On the contrary, the RR-contacts are
energetically penalized.

Our approach can be treated as an extension of
\cite{42}, where only four possible orientations of particles were considered. In this work, the orientation
vector of any molecule can  change continuously, $0<\theta<\pi$ and $0<\phi<2\pi$.
The
Hamiltonian can be written as follows:
\begin{equation}
H =  \sum_{\bf i}(v_\mathbf{i}-\mu)n_{\mathbf{i}}+
 \sum_{nn} u_{\mathbf{ij}} n_\mathbf{i}n_\mathbf{j},
\end{equation}
where $\mu$ is the chemical
 potential. In the above, the first summation is over all lattice sites.  The subscript
 $nn$ means summation over all pairs of the nearest neighbors.

The density is given by
\begin{equation}
 \langle\rho\rangle= {\frac{1}{V}} \sum_{\bf i}n_{\bf i},
\end{equation}
where $V=N^2\, N_z\,$. The potential energy per one lattice
site, $\langle U\rangle $, and the contributions due to external
field, $\langle U_{\text{s}}\rangle $, and due to
particle-particle interactions, $\langle U_{\text{p}}\rangle $ can
be expressed as follows:
\begin{equation}
 \langle U\rangle =\langle U_{\text{s}}\rangle +\langle U_{\text{p}} \rangle = {\frac{1}{V}}\Big( \sum_{\bf i}n_{\bf i}v_{\bf{ i}} +
\sum_{nn}u_{\mathbf{ij}}n_\mathbf{i}n_\mathbf{j} \Big).
\end{equation}

We have carried out Monte Carlo (MC) simulations in the
grand canonical ensemble using the hyperparallel tempering
technique \cite{45,46,47,48}.
 The size
$N_x$  of the system was equal to 48.
Besides the systems exchange between replicas at different
$\mu$ and $T$, a  Monte Carlo step consisted in an attempt to insert a randomly
oriented particle
into the system at a randomly chosen position, an attempt
to remove the existing particle and an attempt to change
the orientation of a selected particle. A number of MC steps
necessary for obtaining solid results depended considerably on the
assumed values of the energy parameters. In the majority of
simulations, we used $10^9$ MC steps (per site) for equilibration
and $10^{10}$ for production runs. However, for certain sets of
energy parameters, the simulations had to be longer.

We have calculated the average density and average potential energy
and corresponding fluctuations, the compressibility
\begin{equation}
 \kappa ={\frac{V}{k_{\text{B}}T}} \left(\langle \rho^2\rangle -\langle \rho\rangle ^2\right),
\end{equation}
and, the heat capacity,
\begin{equation}
 C_V= {\frac{V}{k_{\text{B}} T^2}}\left(\langle H^2\rangle -\langle H\rangle ^2\right).
 \end{equation}

Besides thermodynamic quantities listed above, we have evaluated the histograms of
 $\rho$ and the histograms of $\cos\theta_{i_z}$ in each layer.
The first-order phase transitions were investigated by analysing
the density histograms, and the coexistence of different phases was located using
the equal peak-weight criterion for the density histograms \cite{46}.
Moreover,
we have calculated
 the angular-averaged local densities $\langle \rho_{i_z}\rangle $, the average values
of the cosine of the tilting angle in each layer, $\langle \cos\theta_{i_z}\rangle $, and the
angular-dependent local densities $\rho_{i_z}(\theta,\phi)$.

The energy parameter $\varepsilon$ was taken as unit of energy and
the reduced temperature is $T^*=k_{\text{B}} T/\varepsilon$. Similarly, the
reduced chemical potential is $\mu^*=\mu/\varepsilon$.

\section{Results and discussion}
The aim of this study is to investigate the phase transitions
in the systems. The order-parameter
was the average density. We changed interactions of the
walls with the R-sides of Janus molecules
 from attractive, through neutral, to repulsive, i.e., $t=\varepsilon_{\text{s}}/\varepsilon=5,1,0$ and $-1$ and,
in order to distinguish the systems under study, we abbreviated them
by the label E$t$, where $t=5$, 1, 0 or $-1$.
\begin{figure}[!t]
\begin{minipage}{0.49\linewidth}
\begin{center}
\includegraphics[width=1.0\textwidth]{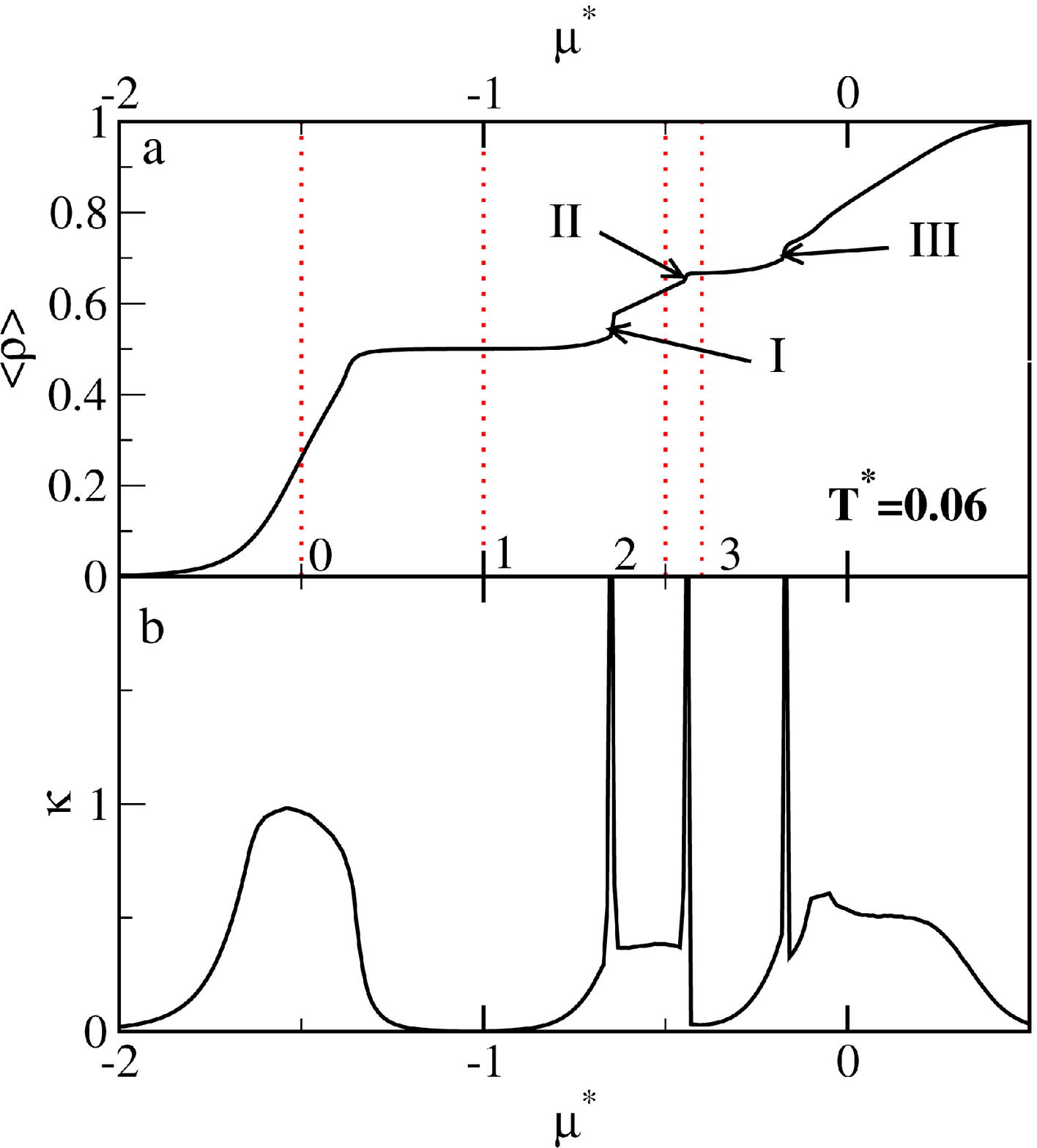}
\end{center}
\end{minipage}
\begin{minipage}{0.49\linewidth}
\vspace{2.37cm}
\begin{center}
\includegraphics[width=1.0\textwidth]{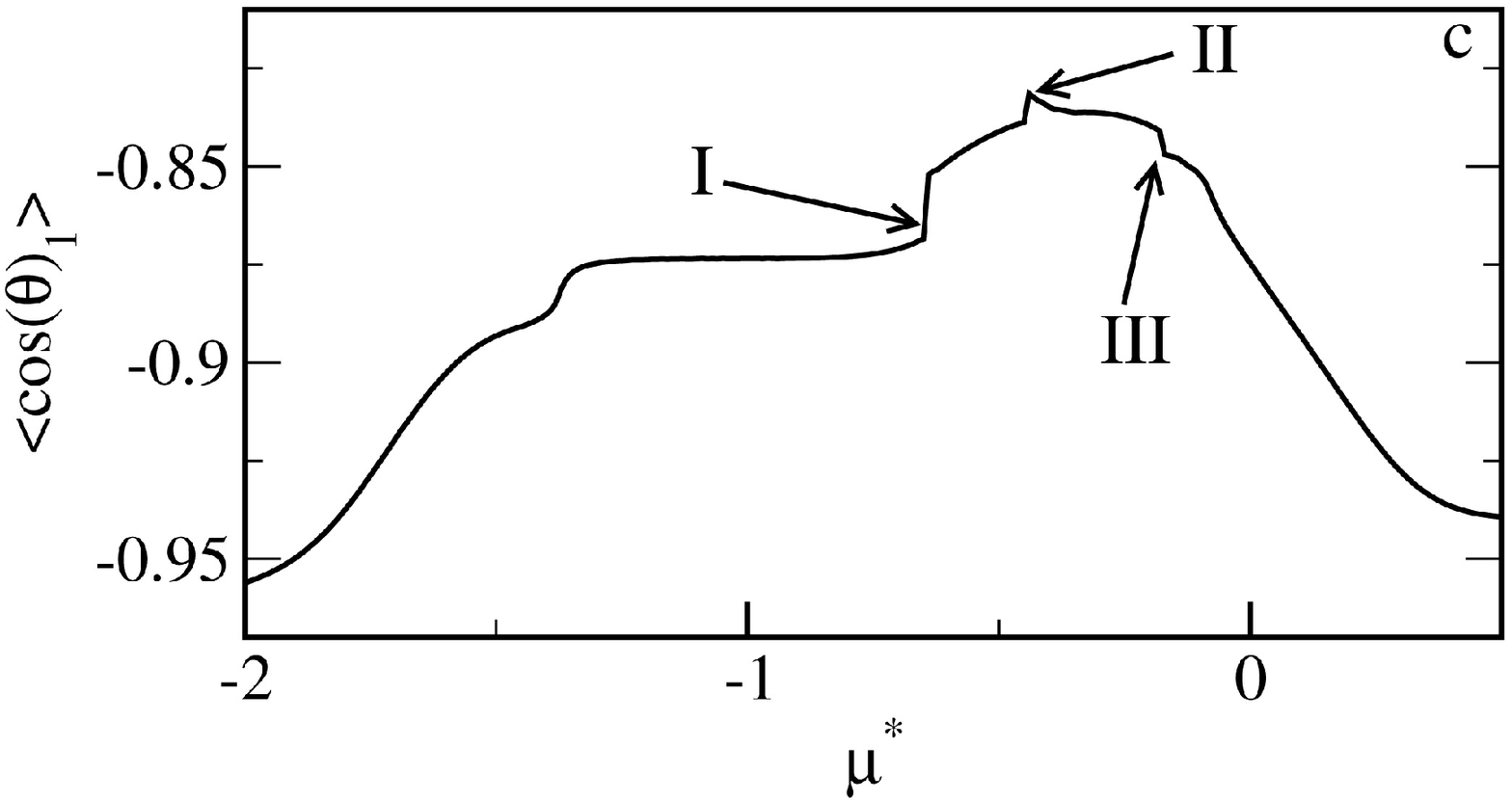}
\end{center}
\end{minipage}
\caption{\label{fig:1}Isotherm [part~(a)], compressibility [part~(b)] and the average tilting angle
of the first-layer molecules [part~(c)] for the system E5. The labels I, II and III
denote the phase transitions. The numbers 0, 1, 2 and 3 and red dotted lines
indicate the states at which the snapshots in figures~\ref{fig:1} and \ref{fig:2} were evaluated. The temperature is
$T^*=0.06$.}
\end{figure}

\begin{figure}[!b]
\begin{center}
\includegraphics[width=0.33\textwidth]{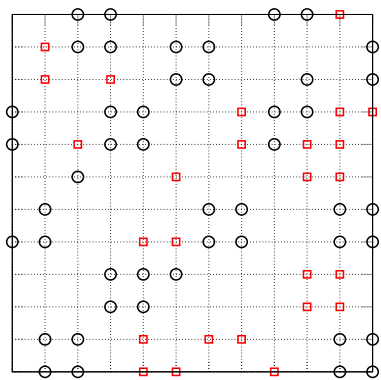} 
\end{center}
\caption{\label{fig:2} (Color online) Snapshot for the state 0 from figure~\ref{fig:1}~(a) for the system E5. Circles
and squares denote the first and the second layer molecules, respectively. For clarity, only 1/16 part
of the simulation cell is shown.}
\end{figure}

\begin{figure}[!b]
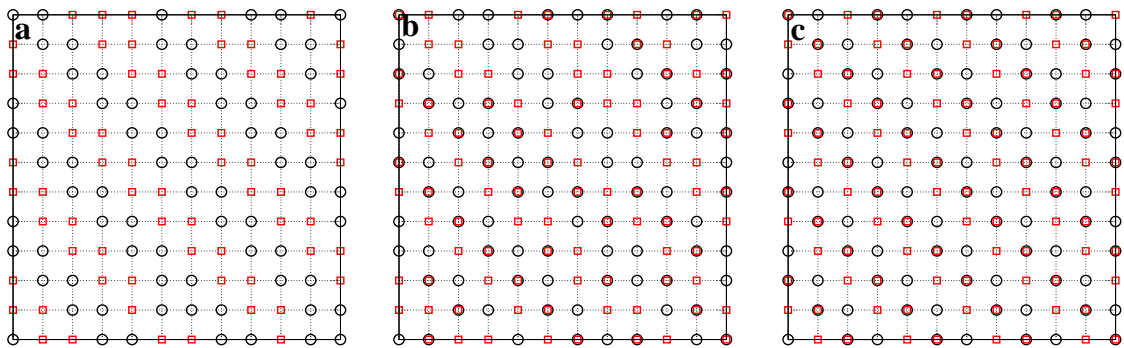

\begin{center}
\includegraphics[width=0.3\textwidth]{F4a.eps} \hspace{0.5cm}
\includegraphics[width=0.3\textwidth]{F4b.eps} \hspace{0.46cm}
\includegraphics[width=0.3\textwidth]{F4c.eps}
\end{center}
\caption{\label{fig:4} (Color online) Snapshots for the states 1, 2 and 3 from figure~\ref{fig:1}~(a)
for the system E5. Circles and squares denote the first and the second layer molecules, respectively. For clarity, only 1/16 part
of the simulation cell is shown.}
\end{figure}

\begin{figure}[!t]
\begin{center}
\includegraphics[width=0.48\textwidth]{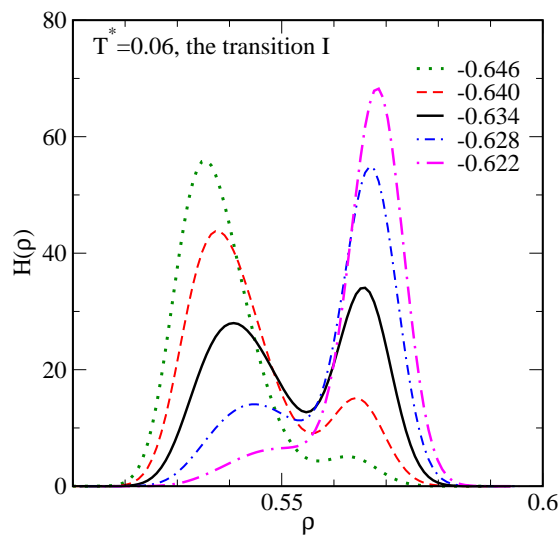} 
\end{center}
\caption{\label{fig:5} (Color online) Unweighted histograms of the density for the transition I in the system E5 at
$T^*=0.06$. The values of the chemical potential are given in the figure.}
\end{figure}

\begin{figure}[!b]
\begin{center}
\includegraphics[width=0.7\textwidth]{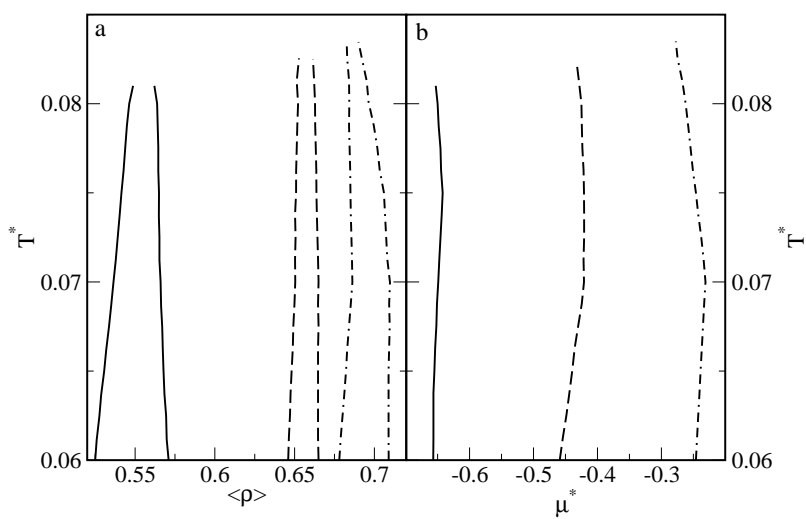} 
\end{center}
\caption{\label{fig:6} Phase diagrams in the density-temperature [part~(a)] and in the chemical potential-temperature
plane for the system E5. The consecutive curves from left to right are
for the transitions I, II and III in figure~\ref{fig:1}~(a).}
\end{figure}

We start the discussion with the case of Janus particles in
relatively strong external field (the system E5). The results are
presented in figures~\ref{fig:1}, \ref{fig:2} and
\ref{fig:4}--\ref{fig:6}. At low values of $\mu$, the molecules
expose their attractive parts towards the pore walls and,
consequently, the molecules in bottom, $i_z=1$, layer expose their
repulsive parts towards molecules  in the second ($i_z=2$) layer.
The external field acts against an ``intrinsic'' tendency to form
an oriented bilayer due to attractive AA-interactions. In
figure~\ref{fig:1}, we show examples of the average density,
$\langle \rho\rangle $ [part~(a)], the values of $\kappa$
[part~(b)] and  the average values of $\langle \cos\theta_1\rangle
$ for the molecules from the bottom
 layer (the values of $\langle \cos\theta_2\rangle $ for  molecules within the top
layer equal to $\langle \cos\theta_2\rangle =-\langle
\cos\theta_1\rangle $). The calculations were carried out at
$T^*=0.06$.

\begin{figure}[!t]
\begin{minipage}{0.46\linewidth}
\begin{center}
\includegraphics[width=0.99\textwidth]{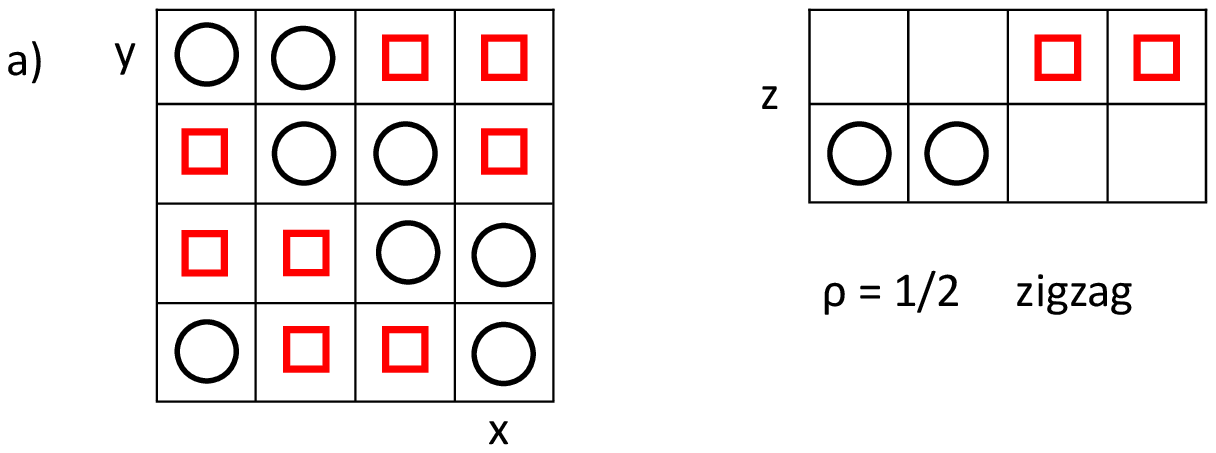}
\end{center}
\end{minipage}
\hspace{0.8cm}
\begin{minipage}{0.46\linewidth}
\vspace{-0.37cm}
\begin{center}
\includegraphics[width=0.99\textwidth]{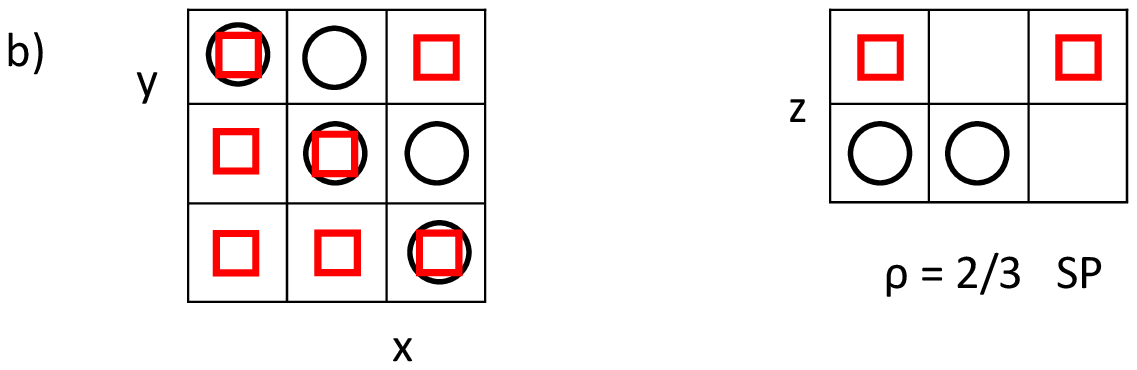}
\end{center}
\end{minipage}
\begin{center}
 \includegraphics[width=0.455\textwidth]{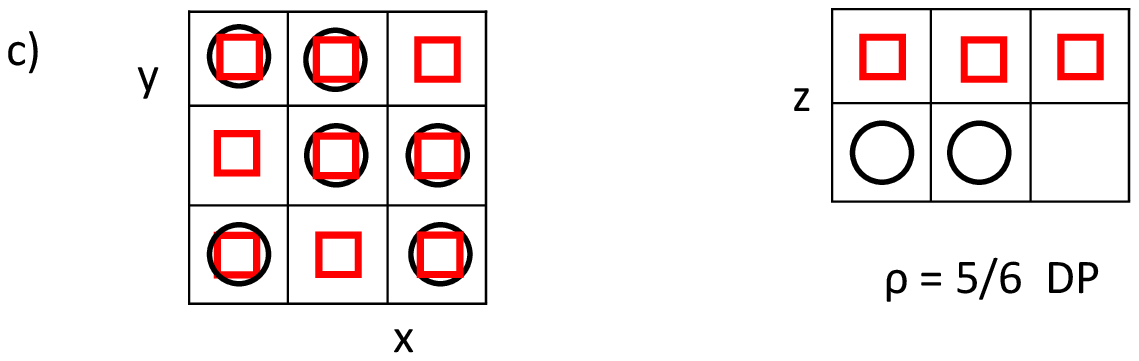}
\end{center}
\caption{\label{fig:3} (Color online) Schematic representation of possible translationally ordered
structures in the system E5: zigzags at $\rho=1/2$~(a), single strips of pillars  (SP) at $\rho=2/3$~(b)
and  double strips of pillars  (DP) at $\rho=5/6$~(c). Circles and squares denote the first and the second
layer molecules, respectively. In left-hand parts, structures formed on the walls are shown.
Right-hand parts illustrate structures perpendicular to the surfaces.}
\end{figure}

Three jumps on the  isotherm in figure~\ref{fig:1}~(a) are seen.
They are marked  as I,  II and III,
respectively. The values
of compressibility, $\kappa$ are discontinuous at these jumps, cf. figure~\ref{fig:1}~(b).
At low values of the chemical potential, $\mu^*$, the molecules
are oriented almost perpendicularly to the surface ($\theta_1\approx\pi$ for
the first and  $\theta_2\approx 0$ for the second
layer molecules). Of course, this orientation results from the strong external
field. For $\mu^*<-1.6$, the formation of clusters
composed of four molecules at each surface is observed (see the snapshot in figure~\ref{fig:2}).
The sites over the cluster formed within the
bottom layer, as well as the sites below the clusters within the upper layer are empty.
In-plane orientation
within each cluster is such that $\langle |\sin(\phi)|\rangle \approx 0.7$, i.e., the values of the
azimuthal angle in each layer for the cluster are, respectively, close to $\pi/4$, $3\pi/4$, $5\pi/4$ and $7\pi$, starting
from the ``low-left'' corner of the cluster.

With an increase of the chemical potential, the clusters start to flow together and
to form
ordered structures. These structures  can be considered as patterns built of the occupied and empty lattice sites.
In such patterns, orientations of particles can change. There is translational order without a more pronounced
orientational order. We have found three translationally ordered phases for $\rho=1/2$, $\rho=2/3$ and $\rho=5/6$.
They are schematically depicted in figure~\ref{fig:3}.
When a half of the lattice sites is occupied, the  zigzag structures on each wall are observed.
There are sloping steplike strips of occupied sites on the surfaces. Similarly to the case of ``squares'',
the zigzags at one layer are in contact with empty sites at the second layer
[see figure~\ref{fig:3}~(a)].
For $\rho=2/3$, strips of pillars are formed. The pillar is built of  particles located ``one atop the other''.
The remaining lattice sites  have empty sites above (or below) the particles. Such a structure is shown in figure~\ref{fig:4}~(b).
In the case of $\rho=5/6$, the zigzag pattern of pillars is formed. In other words, there are double strips of
pillars and single strips of sites with empty sites in one of the layers [cf. figure~\ref{fig:4}~(c)].

We return now to the discussion of phase transitions in the system E5. At the first plateau of
the isotherm [figure~\ref{fig:1}~(a)], the average density  is close to $\langle \rho\rangle \approx 0.5$.
An example of configuration at the plateau of the isotherm ($\mu^*=-1.5$)  is shown in  figure~\ref{fig:4}~(a).
Indeed, one sees here a well pronounced zigzag structure.
Within this region, the average tilting angles of the top (bottom) layers become close to
$\theta_1\approx 5\pi/6$ ($\theta_2\approx \pi/6$),
cf. figure~\ref{fig:1}~(c). At $T^*=0.06$,
the development  of the zigzag structure is a continuous transition.

A further increase of the chemical potential leads to the rearrangement of the molecules
and to the formation of well-developed ``pillars'' between the walls.
The configuration displayed in figure~\ref{fig:4}~(a)  is stable and a significant increase of the
chemical potential is needed in order to enforce more molecules to enter the pore.
The first isotherm jump
is connected with the filling up of the empty zigzag structure. 
This filling leads to the location of the particles
``one atop the other'' and both such particles expose their repulsive parts one towards the other. In order to
reduce the repulsive potential energy, the titling angles for the molecules
located within the first, $\theta_1\,$, and the second layer, $\theta_2\,$, deviate more
from the orientations
perpendicular to the walls. The first step on the isotherm  (I) corresponds to
the transition from the zigzag structure to the structure  containing single strips of pillars (SP).

A typical snapshot of the configuration before the second transition
[marked as II in figure~\ref{fig:1}~(a)] is shown in figure~\ref{fig:4}~(b).
After this transition, the second plateau on the
adsorption isotherm develops. The isotherm plateau  (the average density is $\approx 0.67$) is
shorter than the first one. A representative snapshot of the configuration of molecules
is displayed in figure~\ref{fig:4}~(c). The second
transition is also connected with an increase
of the value of $\cos\theta_1$ to its maximum value. At $T^*=0.06$, the maximum
of $\cos\theta_1$ is close to $-0.83$.
The step II is associated with the transition from the SP-phase to the phase
in which double strips of pillars begin to appear.
At the end of the second plateau, the third step on the adsorption isotherm appears.
We think that the last step (III) is a
transition  from the ordered DP-phase to a phase with the filled empty sites that remained.
This step is connected with a decrease of $\cos\theta_1\,$.

A further
increase of the chemical potential leads to a gradual increase of the density inside the
slit. At high values of the chemical potential, the pore is completely filled. At the third final
plateau, $\cos\theta_1\approx -0.94$. Instantaneously, the azimuthal orientation of molecules
within the filled first and second layers is similar to
that observed at very low values of the chemical potential, when disconnected ``islands'' composed of four
molecules have been formed, i.e., $\langle |\sin(\phi)|\rangle \approx 0.7$.

In figure~\ref{fig:5}, we show examples of unweighted
histograms for the transition I [figure~\ref{fig:1}~(a)].
The distributions exhibit two well
pronounced peaks corresponding to coexisting phases of different densities.
The phase diagrams
for the system E5 are shown in figure~\ref{fig:6}. Part~(a) gives the diagram
in the density-temperature plane, while part~(b) --- in the chemical
potential-temperature plane. The phase diagrams cover a rather narrow
range of temperatures.  The evaluation  of the diagrams at very low temperatures was
difficult and would require extremely long runs to obtain reliable
histograms.  We also did not attempt a precise evaluation
of the critical temperatures.
However, one sees that the critical temperatures of the phase
 transition increase in the order $T_{\text{I}} < T_{\text{II}}< T_{\text{III}}$ [the indices I, II and III
 abbreviate the consecutive transitions from figure~\ref{fig:1}~(a)].
The changes of the density during
phase transition are small and this is a rather striking behavior.
It is also worth stressing that the changes
of the chemical potential with the temperature are non-monotonous.

\begin{figure}[!t]
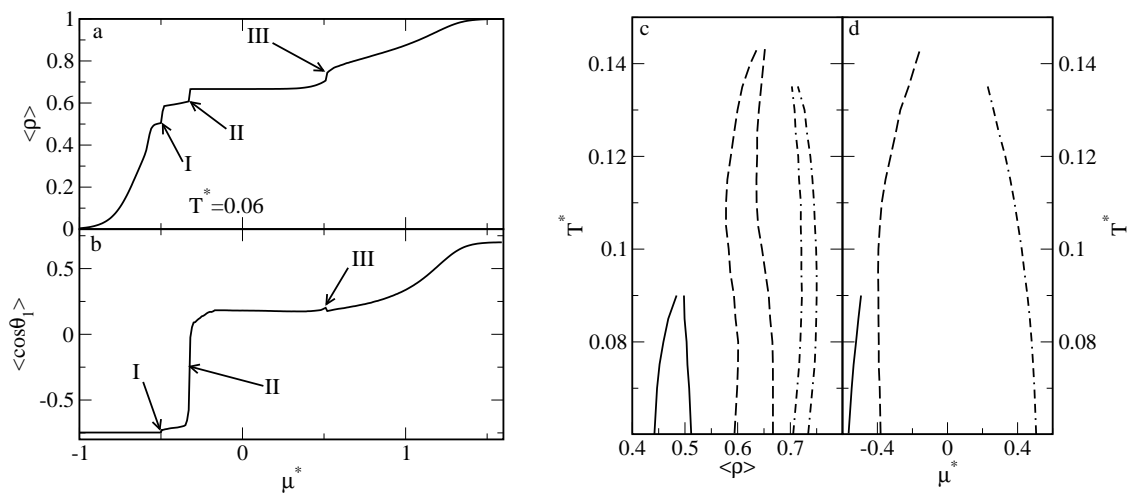

\begin{minipage}{0.45\linewidth}
\begin{center}
\includegraphics[width=0.97\textwidth]{F7ab.eps}
\end{center}
\end{minipage}\hspace{.5cm}
\begin{minipage}{0.5\linewidth}
\begin{center}
\includegraphics[width=1.0\textwidth]{F7cd.eps}
\end{center}
\end{minipage}
\caption{\label{fig:7} Isotherm [part~(a)] and the average tilting
angle of the first-layer molecules [part~(b)] for the system E1.
The labels I, II and III denote the first-order transitions.  The
temperature is $T^*=0.06$. Parts~(c) and (d). Phase diagrams in
the density-temperature [part~(c)] and in the chemical
potential-temperature plane [part~(d)] for the system E1. The
consecutive curves from left to right are for the transitions I,
II and III in part~(a). }
\end{figure}

We now consider the system E1. Similarly to the previous case,
the external field acts against the ``intrinsic'' tendency
to form an oriented bilayer with facing R-sides  of the neighboring Janus particles,
but now this effect is  weaker.
In figure~\ref{fig:7}, we show the adsorption isotherm,
the plot of $\langle \cos\theta_1\rangle $ ($T^*=0.06$) and the phase diagrams for the system E1.

\begin{figure}[!b]
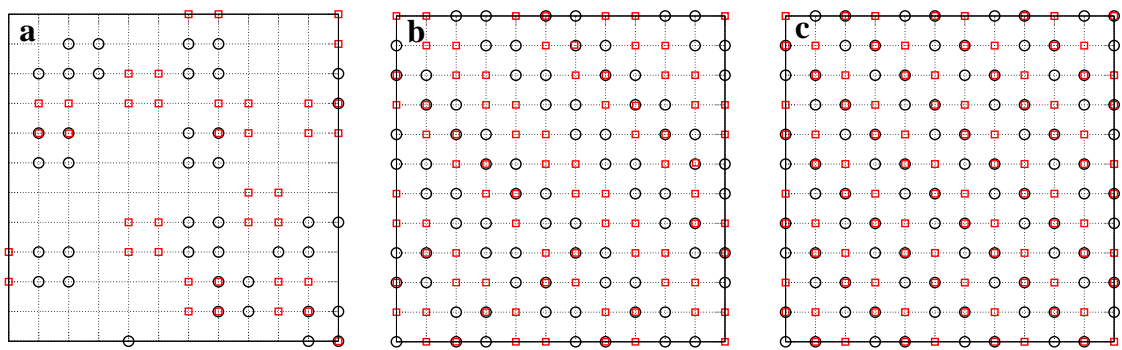

\begin{center}
\includegraphics[width=0.3\textwidth]{F8a.eps} \hspace{0.5cm}
\includegraphics[width=0.3\textwidth]{F8b.eps} \hspace{0.46cm}
\includegraphics[width=0.3\textwidth]{F8c.eps}
\end{center}
\caption{\label{fig:8} (Color online) Snapshots for the chemical potential
$\mu^*=-0.66$ [part~(a)], $-0.49$ [part~(b)] and $-0.13$ [part~(c)]. The calculations
are for the system E1 and at $T^*=0.06$. Circles and squares denote the first and
the second layer molecules, respectively. For clarity, only 1/16 part
of the simulation cell is shown.  }
\end{figure}

One sees three jumps on the adsorption isotherm and
at very low chemical potentials, the structure of the confined fluid is similar
to the system E5. The molecules within each layer form squares.
However, since the external field is weaker, the number of contacts
between the ``square clusters'' in  bottom and in  top layers is  larger than
for the system E5, cf. the snapshot in figure~\ref{fig:8}~(a).
Due to the competition between fluid-fluid and fluid-wall interactions
at low chemical potential, the tilting angle in the system E1 significantly differs
from the tilting angle in the system E5. Now $\langle \cos\theta_1\rangle  \approx -0.74$, i.e.,
$\theta_1$ is close to $3\pi/4$. The in-plane orientation, however, is similar to
the system E5, but the average value $\langle |\sin\phi|\rangle $ is lower, $\langle |\sin\phi|\rangle \approx 0.64$.
This
is the consequence of the existence of contacts between bottom and top layers.

With an increasing chemical potential, a small plateau on the adsorption isotherm
is reached at $\mu^*\approx -0.54$
[figure~\ref{fig:7}~(a)]. Before the transition,
a the zigzag structure is observed.
However, as we have already stressed,  the tilting
angle is different, $\langle \cos\theta_1\rangle  \approx 0.64$ [cf. figure~\ref{fig:7}~(b)]. The transition I
almost does not influence the tilting angle. After the transition, the average density
of the confined fluid is close to 0.6. The translational ordering of the fluid is
developed from
the zigzag structure
in such a way that the stripes at each wall
are ``compressed'' and they alternately contact the filled and empty stripes at the
other wall, cf. figure~\ref{fig:8}~(b).

The second transition [marked as
II in figure~\ref{fig:7}~(a)] leads to the structure
 of the density $\langle \rho\rangle \approx 0.67$ that is shown
in figure~\ref{fig:8}~(c). This transition is connected with
a large change of the tilting angle [figure~\ref{fig:7}~(b)]. At the
isotherm plateau
 that appears for the chemical potential $0.32<\mu<0.45$, the tilting
 angle is $\langle \cos\theta_1\rangle \approx 0.18$.
 This means that the molecules are re-oriented
 from the configuration with molecules facing their attractive parts towards the walls (before the transition)
 to the configuration with the attractive parts of the molecules partially directed towards
 the pore interior (after the transition).
 A small density jump at the transition
 III is not connected with any significant change of the tilting angle.
 For a completely filled pore, the tilting angle $\theta_1$ is close to $\pi/4$.

 Figures \ref{fig:7}~(c) and \ref{fig:7}~(d) show the phase diagram in the density-temperature
 and the chemical potential-temperature planes.
 However, contrary to the system E5, for the system E1
 the critical temperature $T_{\text{II}}$ is the highest and the critical temperature $T_{\text{I}}$ is the lowest.
 Moreover, the critical temperatures for the system E1 are in general higher than for the
 system E5, which is the consequence of
 the total potential energy  being considerably lower in the case of the system E5.

\begin{figure}[!b]
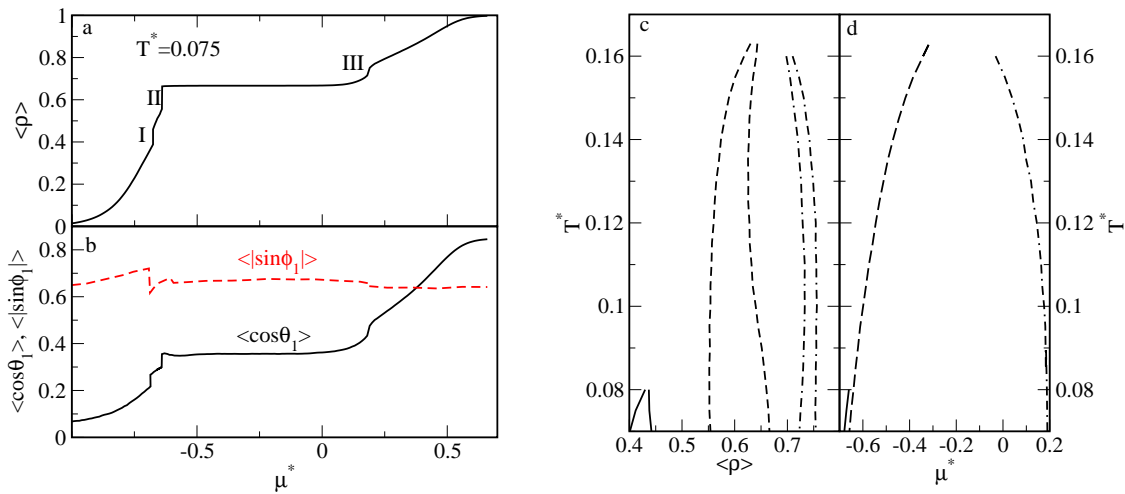

\begin{minipage}{0.45\linewidth}
\begin{center}
\includegraphics[width=0.96\textwidth]{F9ab.eps}
\end{center}
\end{minipage}\hspace{.5cm}
\begin{minipage}{0.5\linewidth}
\begin{center}
\includegraphics[width=1.0\textwidth]{F9cd.eps}
\end{center}
\end{minipage}
\caption{\label{fig:9} (Color online) Isotherm [part~(a)], the average tilting angle and the values of  $\langle |\sin\phi|\rangle $
of the first-layer molecules [part~(b)] for the system E0.
The labels I, II and III
denote the first-order transitions.  The temperature is
$T^*=0.075$. Parts~(c) and (d).
Phase diagrams in the density-temperature [part~(c)] and in the chemical potential-temperature
plane [part~(d)] for the system E0. The consecutive curves from left to right are
for the transitions I, II and III shown in part~(a).}
\end{figure}

\begin{figure}[!t]
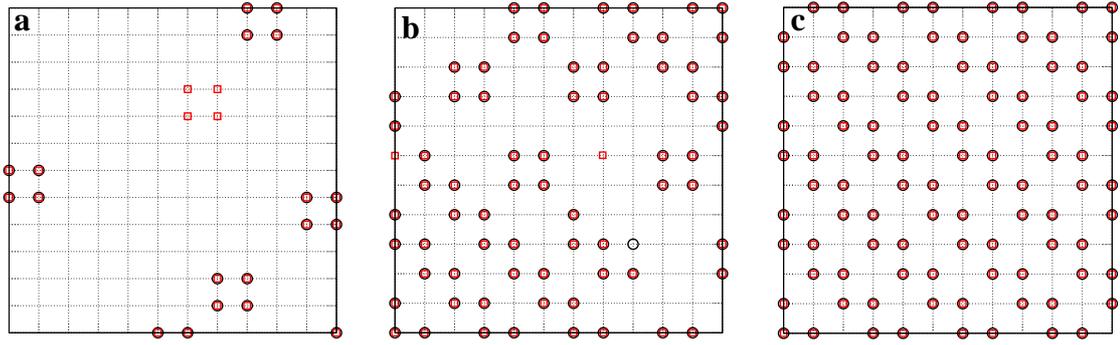

\begin{center}
\includegraphics[width=0.3\textwidth]{F10a.eps} \hspace{0.5cm}
\includegraphics[width=0.3\textwidth]{F10b.eps} \hspace{0.46cm}
\includegraphics[width=0.3\textwidth]{F10c.eps}
\end{center}
\caption{\label{fig:10} (Color online) Snapshots for the chemical potential
$\mu^*=-0.8$ [part~(a)], $-0.66$ [part~(b)] and $-0.2$ [part~(c)]. The calculations
are for the system E0 and at $T^*=0.075$. Circles and squares denote the
first and the second layer molecules, respectively. For clarity, only 1/16 part
of the simulation cell is shown.  }
\end{figure}

 We next discuss the system E0, where the pore walls are just hard walls and  orientation
 effects result entirely from the fluid-fluid interactions. Figure \ref{fig:9} shows the isotherm [part~(a)],
 tilting angles [part~(b)] and the phase diagrams [parts~(c) and (d)]. Additionally,
 in part~(b) we have also displayed the values of
 $\langle |\sin\phi|\rangle $.  As in the previous cases, three first-order phase transitions
 appear within the investigated range of temperatures.
 However, we observe the formation of considerably different
  ordered structures in the system. Before the
 first transition, the confined molecules form ``pillars'' built of 8 molecules, four at
 each wall, cf. figure~\ref{fig:10}~(a).  We also observe the appearance  of ``incomplete pillars'', but
 the analysis of the distribution of the cluster sizes indicated that
 for the densities $0.05< \langle\rho \rangle < 0.2$, up to 80-85\% of the clusters are built
 of 8 molecules.
 It should be stressed that now the particles are differently oriented with respect to the walls.
 At very low densities, the tilting angle $\theta_1$ is close to $85^\circ$,
 i.e., the orientation of the molecules is almost perpendicular to the vector normal to
 the surface. With an increase of $\langle \rho\rangle $ to $\approx 0.2$,
 the tilting angle slightly decreases to $82^\circ$. Instantaneously, the azimuthal angle is
 characterized by $\langle |\sin\phi|\rangle  \approx 0.69$, i.e., the principal azimuthal angle is close to $\pi/4$.

\begin{figure}[!b]
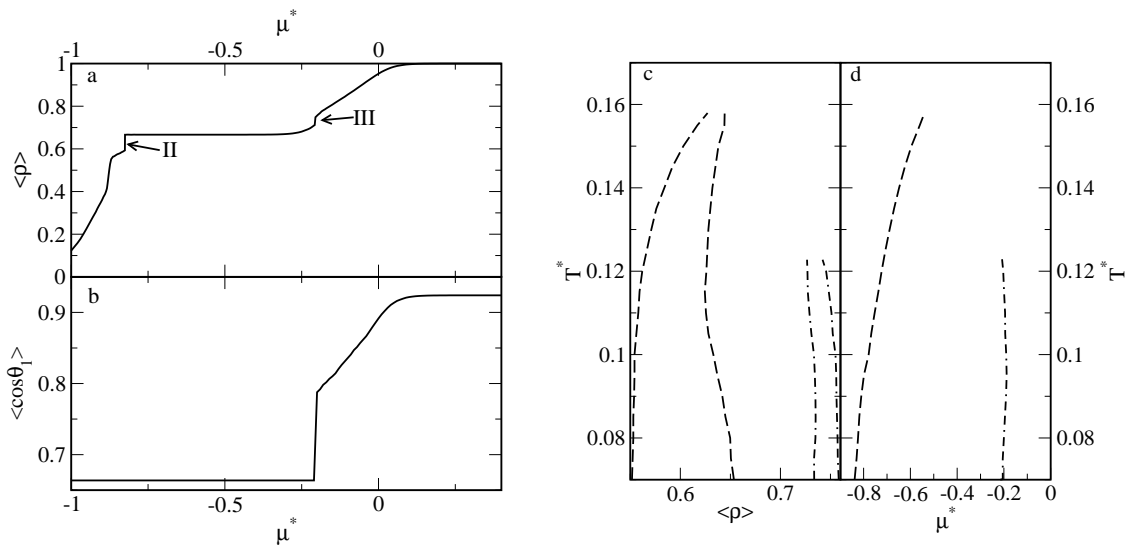

\begin{minipage}{0.45\linewidth}
\begin{center}
\includegraphics[width=0.97\textwidth]{F11ab.eps}
\end{center}
\end{minipage}\hspace{.5cm}
\begin{minipage}{0.5\linewidth}
\vspace{0.55cm}
\begin{center}
\includegraphics[width=1.0\textwidth]{F11cd.eps}
\end{center}
\end{minipage}
\caption{\label{fig:11}  Parts~(a) and (b) show the adsorption
isotherm and the values of $\langle \cos\theta_1\rangle $,
respectively at $T^*=0.075$, while parts~(c) and (d) --- the phase
diagrams in the density-temperature [part~(c)] and in the chemical
potential-temperature [part~(d)] plane for the system $\textrm{E}-1$.  In
parts~(c) and (d) the consecutive curves  from left to right are
for the transitions  II and III in part~(a).}
\end{figure}

 With an increase of $\langle \rho\rangle $ ($\mu$), the zigzag structures are formed on both walls.
 In contrast to the systems E5 and E1, the zigzags are connected.
The layers at the top wall are built over the layers at the bottom wall.
  After the transition,
 the tilting angle is close to $\theta_1\approx 0.75^\circ$. The first
 transition also leads  to a small jump of $\langle |\sin\phi|\rangle $.\

 The second transition (II) is connected
 with the development of the structure shown in figure~\ref{fig:10}~(c).
 After this transition, the density is close
 to $\langle\rho \rangle \approx 0.67$, the tilting angle
 is close to $\theta_1\approx 70^\circ$ and $\langle |\sin\phi|\rangle  \approx 0.67$.
  Also, during the third transition,
 the changes of $\langle \cos\theta_1\rangle $ and of $\langle |\sin\phi|\rangle$
 are very small.
 After the transition III, further filling the pore is
 continuous and, finally, for a completely filled pore, $\langle \cos\theta_1\rangle \approx 0.84 $, i.e.,
 $\theta_1 \approx 32^\circ$.  For  the entire range of densities, the changes
 of the azimuthal angle are small:  $0.61<\langle|\sin\phi|\rangle<0.7$.

 In figures \ref{fig:9}~(c) and \ref{fig:9}~(d) we show the phase diagrams for the system E0. The first transition
 appears only at low temperatures and its critical
 temperature is lower than for the system E1.  The
 critical temperatures of the second and third transitions, however, are slightly
 higher than for the system E1.

 Figure \ref{fig:11} summarizes the results obtained for the system E-1 at temperature $T^*=0.075$.
 In contrast to the previous cases, only two first-order
 transitions occur within the investigated range of temperatures.
 The ``survived'' transitions correspond to the transitions II and III
 observed in the systems investigated previously.
 Indeed, the
 the first jump
 on the  isotherm
 is connected with the formation of the stripes. After
 the transition,  $\langle \rho\rangle  \approx 0.67$ and the translational
 structure of the system is the same as
 displayed in figure~\ref{fig:10}~(c). This transition (we still abbreviate it as II) is not
 connected with any pronounced
 change of the tilting angle and $\langle \cos\theta_1\rangle \approx 0.67$, both
 before and after the transition. However, a big change of the tilting
 angle occurs during the second transition [figure~\ref{fig:11}~(c)]. For a completely filled pore,
 at low temperatures,
 $\langle \cos\theta_1\rangle \approx 0.92$, i.e., the particles located at the bottom and top walls
 expose their attractive parts one towards the other.

\section{Summary}

We have carried out Monte Carlo simulations
of Janus-like particles on a cubic lattice, confined in very narrow slits. The particles
consisted of two hemispheres that mutually interacted via  short-range
directional potential. Directional interactions
were also assumed between  particles and pore walls. We formulated a simple model describing
these interactions.

A great majority of the calculations have been performed for
bilayers. In this case, we have studied the effects of the competition between
fluid-fluid and fluid-pore walls interactions. The principal aim of our
calculations was to investigate the phase transitions in the system. The
transitions were located by analyzing the histograms of the density.
The behavior of the histograms and of other thermodynamic quantities
(the heat capacities and compressibilities) suggests the first-order
character of the transitions. However,  our calculations were carried out
for one (though rather large) size of the system in $X$ and $Y$ directions, and
we did not perform the analysis of the dependence of the calculated
characteristics on the systems size. To be precise, one should
stress that only the analysis of the finite size effects \cite{43aa,44aa} could
conclusively prove the first-order character of the observed transitions.
Nevertheless, we believe that our simulations correctly
reveal the number and the range of temperatures, densities and
chemical potential of the first-order transitions
in the systems under study.

The oriented structures found for the
systems in which the fluid-wall interactions are strong and act contrary to the
fluid-fluid interactions differ from those appearing in the systems with
neutral walls or with the walls attracting the repulsive parts of fluid molecules.
In the former case, at low values of the chemical potential, the structures
(squares or zigzags) formed at one wall connect the empty sites
at the other wall. The extortion of overlaps of the structure
at both walls requires a significant increase of the chemical potential.
This is particularly visible for the system E5. In the latter case of
fluid-wall interactions, the tendency to form overlapping structures is
enforced by reinforcement of the orientational effects resulting from fluid-fluid
interactions by fluid-wall forces.
In all the studied bilayers, the single and double strips of pillars were formed at higher values of the chemical potential.

One can expect that depending on the relations between parameters characterizing  Janus-like mole\-cu\-les
confined in wider  pores,  various complicated ordered structures can appear. This opens up new directions
in the modelling of modern materials. We hope that
the presented results provide a guidance for a further study of
concrete systems within the framework of more realistic models.

\newpage

\ukrainianpart
\title
{Фазовий перехід першого роду в граткових двошарах Янус-подібних частинок: Монте Карло симуляції}

\author{M.~Борувко, A.~Патрикєєв, В. Жиско, С.~Соколовскі}
\address{
Відділ моделювання фізико-хімічних процесів,
Університет Марії Кюрі-Склодовської, Люблін, Польща}

\makeukrtitle

\begin{abstract}
Досліджуються фазові переходи першого роду в граткових двошарах Янус-подібних частинок, обмежених щілиноподібними порами.
Розглядається кубічна гратка з молекулами, які можуть вільно змінювати свою орієнтацію на вузлі гратки. Крім того, молекули можуть взаємодіяти
зі стінками пори за рахунок сил, залежних від орієнтації. Проведені обчислення обмежуються випадками двошарів.  Наголос зроблено на конкуренції взаємодій плин-стінка та плин-плин.
Орієнтовані структури, сформовані у системах, в яких взаємодії плин-стінка, що діють протилежно до взаємодій плин-плин, відрізняються від структур, що з'являються в системах з нейтральними стінками чи зі стінками, які притягують відштовхувальні частини молекул плину.

\keywords Монте Карло симуляції, граткові Янус-частинки, фазові переходи, двошари

\end{abstract}

\end{document}